\documentclass[sigconf]{acmart}
\AtBeginDocument{%
  \providecommand\BibTeX{{%
    \normalfont B\kern-0.5em{\scshape i\kern-0.25em b}\kern-0.8em\TeX}}}

\copyrightyear{2024}
\acmYear{2024}
\setcopyright{acmlicensed}\acmConference[CHI '24]{Proceedings of the CHI Conference on Human Factors in Computing Systems}{May 11--16, 2024}{Honolulu, HI, USA}
\acmBooktitle{Proceedings of the CHI Conference on Human Factors in Computing Systems (CHI '24), May 11--16, 2024, Honolulu, HI, USA}
\acmDOI{10.1145/3613904.3642482}
\acmISBN{979-8-4007-0330-0/24/05}

\usepackage{graphicx}
\usepackage{caption}
\usepackage{subcaption}

\begin{document}

\title[Evaluating the Experience of LGBTQ+ People Using LLM-Based Chatbots for Mental Health Support]{Evaluating the Experience of LGBTQ+ People Using Large Language Model Based Chatbots for Mental Health Support}

\author{Zilin Ma}
\authornote{Equal contributions}
\email{zilinma@g.harvard.edu}
\orcid{0000-0002-7259-9353}
\affiliation{%
  \institution{Intelligent Interactive Systems Group\\Harvard School of Engineering and Applied Sciences}
    \streetaddress{150 Western Ave.}
  \city{Allston}
  \state{MA}
  \country{USA}
  \postcode{02134}
}

\author{Yiyang Mei}
\authornotemark[1]
\email{yiyang.mei@emory.edu}
\orcid{0009-0001-2923-0729}
\affiliation{%
  \institution{Law School\\Emory University}
    \streetaddress{1301 Clifton Road}
  \city{Atlanta}
  \state{GA}
  \country{USA}
  \postcode{30322}
}

\author{Yinru Long}
\email{yinru.long@vanderbilt.edu}
\orcid{0000-0003-4830-9894}
\affiliation{%
  \institution{Psychology and Human Development\\Peabody College\\Vanderbilt University}
    \streetaddress{230 Appleton Pl \#5721}
  \city{Nashville}
  \state{TN}
  \country{USA}
  \postcode{37203}
}

\author{Zhaoyuan Su}
\email{nick.su@uci.edu}
\orcid{0000-0002-5647-8439}
\affiliation{%
  \institution{Donald Bren School of Information and Computer Sciences\\University of California Irvine}
    \streetaddress{204 Aldrich Hall Irvine}
  \city{Irvine}
  \state{CA}
  \country{USA}
  \postcode{90293}
}

\author{Krzysztof Z. Gajos}
\email{kgajos@eecs.harvard.edu}
\orcid{0000-0002-1897-9048}
\affiliation{%
  \institution{Intelligent Interactive Systems Group\\Harvard School of Engineering and Applied Sciences}
  \streetaddress{150 Western Ave.}
  \city{Allston}
  \state{MA}
  \country{USA}
  \postcode{02134}
}

\renewcommand{\shortauthors}{Ma and Mei, et al.}

\begin{abstract}
LGBTQ+ individuals are increasingly turning to chatbots powered by large language models (LLMs) to meet their mental health needs. However, little research has explored whether these chatbots can adequately and safely provide tailored support for this demographic. We interviewed 18 LGBTQ+ and 13 non-LGBTQ+ participants about their experiences with LLM-based chatbots for mental health needs. LGBTQ+ participants relied on these chatbots for mental health support, likely due to an absence of support in real life. Notably, while LLMs offer prompt support, they frequently fall short in grasping the nuances of LGBTQ-specific challenges. Although fine-tuning LLMs to address LGBTQ+ needs can be a step in the right direction, it isn't the panacea. The deeper issue is entrenched in societal discrimination. Consequently, we call on future researchers and designers to look beyond mere technical refinements and advocate for holistic strategies that confront and counteract the societal biases burdening the LGBTQ+ community. 
\end{abstract}




\received{14 September 2023}
\received[revised]{12 December 2023}
\received[accepted]{19 January 2024}

\maketitle
\section{Introduction}
The increase in social isolation coupled with inadequate access to professional mental health services has led many to turn to large language model (LLM) based chatbots in hopes of finding connection and support for their mental wellbeing. Platforms like ChatGPT, Replika, Anima, Kajiwoto, and Character AI have gained immense popularity, with millions using them for immediate, discreet social and emotional support ~\cite{metz2020riding}. These LLM-based companions provide comfort to those feeling lonely or in difficult situations by offering conversational engagement anytime and anywhere~\cite{ma_understanding_2023, metz2020riding}. The advanced linguistic capabilities of LLM-based chatbots offer users more context-aware and responsive interactions, distinguishing them from the earlier pre-LLM chatbots~\cite{kasneci_chatgpt_2023}.

The potential of LLM-based chatbots is most striking when considering their impact on historically marginalized communities like the LGBTQ+ (lesbian, gay, bisexual, transgender, queer, and/or questioning)~\cite{Henkel_2023_utaut}. LGBTQ+ individuals face significantly higher rates of depression (57\%), anxiety (70\%), and suicidal ideation (41\%) compared to their heterosexual cis-gendered peers~\cite{veale_enacted_2017, meyer_prejudice_2003}. Beyond these alarming statistics, LGBTQ+ people also navigate a daily landscape marred by discrimination, bullying, and stigma tied to their gender and sexual identities, and endure a glaring absence of representation in the mainstream culture~\cite{meyer_prejudice_2003}. This lack of representation and systemic marginalization deter them from seeking professional therapeutic assistance, especially when there is a risk of encountering non-affirmative therapists~\cite{meyer_prejudice_2003,drescher_issues_2020}.

Although LLM-based chatbots seem to offer a valuable and inclusive mental health resource for the LGBTQ+ community, potentially bridging gaps in traditional therapy accessibility~\cite{fitzpatrick_delivering_2017, ma_understanding_2023}, their deployment raises substantial concerns. Biases embedded in these chatbots can perpetuate harmful stereotypes. LGBTQ+ users, who are often underrepresented in the training datasets, can encounter unintentional reinforcement of damaging narratives with regard to their identities~\cite{felkner_winoqueer_2023, stochastic-parrot}. Furthermore, as people's reliance on these platforms increases~\cite{ma_understanding_2023}, there are growing apprehensions regarding the chatbots' ability to truly understand the nuances of LGBTQ+ identities and the depth of human emotions~\cite{Wang2021AnEO, felkner_winoqueer_2023}. Consequently, to investigate these challenges, we ask: 

\begin{itemize}
    \item What benefits can LLM-based chatbots provide to LGBTQ+ people in terms of mental wellness support?
    \item Do LGBTQ+ people have additional purposes of use for LLM-based chatbots compared to non-LGBTQ+ people?
    \item Can LLM-based chatbots meet LGBTQ+ people's mental wellness needs regarding their identity?
\end{itemize}

We interviewed 31 participants (18 identifying as LGBTQ+ and 13 as non-LGBTQ+) about their usage of LLM-based chatbots for mental wellness support. We specifically asked the LGBTQ+ participants how LLM-based chatbots supported their mental wellness needs regarding their LGBTQ+ identity. We had the following findings:
\begin{itemize}
    \item For both LGBTQ+ and non-LGBTQ+ participants, LLM-based chatbots offer immediate support and accessibility, create a safe environment for intimate conversations, foster strong emotional bonds between the chatbots and the users, and are useful for developing social skills.
    \item For both LGBTQ+ and non-LGBTQ+ participants, the ease of usage and emotional bonding has the potential to encourage adherence to therapy regimens when applied in mental health, but also risk over-reliance.
    \item LGBTQ+ participants use chatbots due to a lack of real-life support, seeking guidance on topics like coping with discrimination or seeking identity affirmation.
    \item LGBTQ+ participants use LLM-based chatbots to rehearse LGBTQ+-specific experiences such as coming out and dating as an LGBTQ+ person.  
    \item LLM-based chatbots cannot \textit{completely} address the nuances in the emotional needs of LGBTQ+ people due to their overly generalized responses.
    \item LLM-based chatbots offer suggestions that might be ignorant of the ever-changing societal norms (e.g., coming out to unsupportive parents), such that if the users fully follow the advice, they risk danger to themselves.
\end{itemize}

Our results show that LLM-based chatbots have a long way to go before they can fully address the needs of LGBTQ+ people's mental health needs. Moreover, because we identified that the main motivation for using LLM-based chatbots for mental health was the lack of social support, we argue that designing solutions that address the societal stigma against LGBTQ+ people should be prioritized over optimizing LLMs on LGBTQ+ people's needs. Therefore, we recommend ways to improve LLMs for the specific use cases of LGBTQ+ people, and also possible socio-technical solutions to address stigmas LGBTQ+ people face online.


\section{Related Work}
This section references societal norms, behaviors, and attitudes found within contemporary Western cultures. It's essential to note that the literature summarized here may not necessarily reflect or encompass the nuances and perspectives of Asian, African, Latin American, or even Eastern European cultures. 

\subsection{LGBTQ+ People's Online Experiences }
Online technologies offer significant benefits to LGBTQ+ individuals, especially those who lack real-life support from family or friends~\cite{Henkel_2023_utaut, FOX2016635, troiden1988gay, mckenna_coming_1998, dehaan2013interplay}. These platforms provide crucial access to interpersonal and systemic resources, as shown by the success of initiatives like The Trevor Project. Founded to prevent suicide and offer crisis intervention, The Trevor Project has amassed over 2 million followers on platforms like X and Instagram~\cite{trevorproject2023}. Similarly, social media networks like TikTok and Tumblr have become vital spaces for LGBTQ+ individuals to explore and express their sexual orientation and gender identity~\cite{simpson_2021_for_you, devito_2021_values}. In other cases, online technologies help LGBTQ+ people to navigate identity-related challenges, engage with supportive communities, and access educational resources about LGBTQ+ issues~\cite{mcinroy_platforms_2019}. These online technologies are crucial to LGBTQ+ people, as they continue to experience disproportionate risks and limited access to support offline, including at home, at school and in their communities~\cite{Henkel_2023_utaut, FOX2016635, craig_you_2014, mcinroy_platforms_2019}. 


However, online technologies can sometimes fall short of meeting the needs of the LGBTQ+ community, as they do not center LGBTQ+ people in the design process~\cite{Haimson_2020_trans}. For example, Tumblr's 2018 ban on ``adult content'' disproportionately affected transgender users~\cite{oliver_2021_tumblr}. Many transition-related posts were mistakenly categorized as adult material, inadvertently marginalizing this group. Similarly, YouTube's policy of labeling LGBTQ+ content as ``adult'' has further isolated these communities~\cite{alkhatib_street-level_2019}. Facebook's insistence on real names fails to recognize the value of anonymity for LGBTQ+ individuals, which is indispensable for their safety and freedom~\cite{boyd2007social}. To optimize monetization, many content creators, mostly non-LGBTQ+ members, sometimes resort to tactics like ``queerbaiting''~\cite{moolenijzer_2023_queerbaiting}. Queerbaiting is a marketing technique used to attract the LGBTQ+ audience by hinting at same-sex relationships or LGBTQ+ themes without actually depicting or confirming them. This tactic is often criticized for exploiting LGBTQ+ themes for commercial gain without providing meaningful representation~\cite{moolenijzer_2023_queerbaiting}.  Dating websites, while providing a means of connection for individuals, still frequently perpetuate racism and ableism, excluding marginalized groups within the LGBTQ+ community, such as queer people of color and those living with HIV~\cite{Hutson2018, ma_2022_dating, liang_surveillance_2020}. Additionally, the disproportionate prevalence of cyberbullying against queer individuals compared to their heterosexual counterparts highlights the significant challenges faced in online spaces by the LGBTQ+ community~\cite{robyn_2012_cyberbullying}. 

\subsection{Digital Mental Support Technology for LGBTQ+ Individuals}
The LGBTQ+ community experiences greater mental health challenges such as higher levels of depressive symptoms, engaging in more non-suicidal-self-injury, and having more suicidal thoughts and behaviors compared with heterosexual, cisgender peers~\cite{amos_mental_2020, irish_depression_2019,toomey_coping_2018, semlyen_sexual_2016, veale_enacted_2017, trevorproject2023}. The stress of coming out also lead to increased depressive and anxiety symptoms and suicidal ideation~\cite{ryan_family_2010, cox_stress-related_2010, hilton_family_2011, meyer_prejudice_2003}. Minority stress theory suggests that structural stigma against LGBTQ+ people, interpersonal discrimination, and internalized stigma all exacerbate the mental health challenges of this population, resulting in feelings of alienation and distress~\cite{meyer_minority_1995, cox_stress-related_2010, herdt_introduction_1989}. Additionally, the frequent dismissal of LGBTQ+ youth experiences as mere ``teenage angst''~\cite{ryan_family_2009} contributes to a sense of disconnection and isolation, inflicting feelings of being unloved or misunderstood within their support systems, or even more severe consequences such as homelessness~\cite{robinson_conditional_2018}. Social support from family and friends is crucial for LGBTQ+ individuals to mitigate stress~\cite{clark2023adolescents}. However, LGBTQ+ people often report less perceived family support than their heterosexual, cisgender peers and face challenges in peer relationships~\cite{saewyc2011research}. 

Given the dismissal of their concerns and the lack of availability of LGBTQ+-specific mental health care, digital therapies, such as those involving digital cognitive behavioral therapy (dCBT), have shown promise as an alternative mental health support avenue for LGBTQ+ individuals. By providing self-guided, affordable, accessible, and private mental health care, they address key barriers to traditional therapy, including long waiting times, extended treatment duration, and traveling costs~\cite{mcdermott_queer_2016, hollis_annual_2017, andersson_advantages_2014}. Nonetheless, digital therapies like dCBT demand a significant amount of commitment and self-monitoring~\cite{espie_use_2013, beun_persuasive_2013}. Other limitations such as low adherence rate, technical difficulties and sophistication, and privacy concerns significantly hinder effectiveness~\cite{espie_use_2013, beun_persuasive_2013}. 

In addition to the online delivery of mental health services, digital communities, especially those fostered on social media and associated with LGBTQ+ organizations, have emerged as pivotal spaces supporting LGBTQ+ mental well-being~\cite{lucassen_how_2018}. They are frequently used by LGBTQ+ youth, providing emotional sustenance, guidance, and a sense of belonging~\cite{lucassen_how_2018, mcdermott_social_2018}. In addition, they also offer a safe milieu for self-expression and identity exploration, creating an oasis where shared experiences and mutual understanding can bring solace~\cite{craig_you_2014}. 

Online platforms offer advice and guidance on societal challenges ranging from addressing discrimination to identifying LGBTQ+-friendly resources. This function is especially crucial for individuals lacking access to LGBTQ+ resources in real life or a supportive and intimate environment~\cite{ryan_family_2009}. Furthermore, these platforms ameliorate feelings of isolation that are prevalent among LGBTQ+ youth, particularly for those who are still in the closet or are in less accepting environments. Websites such as The Trevor Project and platforms like LGBTQ+ forums on Reddit or specialized apps like TrevorSpace~\cite{TrevorSpace2023} have become sanctuaries for many LGBTQ+ youths. These spaces provide them with an opportunity to share their stories, listen to the experiences of others, and realize they're not alone. Such platforms often have features like chat services, community boards, and resources specifically tailored to provide peer support and information. While online platforms offer valuable social support, it is important to note that they are not a substitute for professional mental health services. These online platforms can have varied content quality and have the potential to expose users to cyberbullying or negative comparisons due to a less strict code for data privacy and protection mechanisms compared to working with a therapist~\cite{coulson_pros_2016, auriemma_cyberbullying_2020}. 

\subsection{Mental Wellness Chatbots} 
\subsubsection{Pre-LLM chatbots for mental wellness}

Before the emergence of LLMs, chatbot architecture primarily consisted of three approaches: rule-based, retrieval-based, and a combination of both~\cite{Wang2019, xiao2020, Deshpande2017ASO}. Rule-based chatbots operate on predefined rules, linking user inputs to specific responses~\cite{Eliza}. Retrieval-based chatbots used machine learning algorithms to choose responses from a preset database according to user inputs~\cite{kadlec_improved_2015, Lowe2017TrainingED}. There were also generative systems that were built on neural network architecture like Sequence-to-Sequence (Seq2Seq) models~\cite{serban_multiresolution_2017, serban_building_2016, wu-etal-2017-sequential,sutskever_sequence_2014}. Although capable of generating unique responses, these models were limited by the need for extensive training data, significant computational power, and the challenge of maintaining context in long conversations~\cite{yan_emotionally_2022, li-etal-2016-deep, scotti2023, huang2021}. Pre-LLM chatbots offered high control (to the creators) due to their structured design. Their accessibility and instant response features made pre-LLM chatbots popular in mental health applications. Research indicats that mental health chatbots have had positive impacts in reducing symptoms of depression and anxiety, and enhancing therapeutic alliance, acceptability, and likeability, particularly during the COVID-19 pandemic~\cite{he2022mental,ollier2023can, abd-alrazaq_effectiveness_2020, abd-alrazaq_overview_2019, su2021casocialdistancing}. 

Despite the initial successes and widespread use of pre-LLM chatbots in mental health applications, as evidenced by numerous studies on their acceptability and usability, there remains a significant gap in research specifically addressing their effectiveness in improving mental health outcomes. This lack of comprehensive research presents a challenge in fully understanding and evaluating the impact of these chatbots in mental wellness care~\cite{boucher_artificially_2021,fitzpatrick_delivering_2017, darcy_anatomy_2022, durden_changes_2023, boucher_artificially_2021, su_analyzing_2020}. Prior research highlights that the success of mental wellness chatbots largely depends on sociotechnical aspects and therapeutic relationships~\cite{liao2021design}. Pre-LLM chatbots, given their technological limitations, often struggle to effectively address these crucial elements. Significant drawbacks, such as limited linguistic or contextual understanding, often led to unnatural or irrelevant conversations, reducing users' willingness to engage with these chatbots, making interactions less convincing and supportive, and potentially limiting therapeutic benefits~\cite{chatbot_review, lacking_linguistics_ability, chatbot_emotion, Pizzi_2023_chatbot}. Furthermore, these chatbots struggled to adapt and learn from user information, failing to cater to individual needs~\cite{chatbot_personalization}. Consequently, chatbots frequently fall short of genuinely understanding and responding to emotional nuances. This issue is particularly pronounced among marginalized communities, such as LGBTQ+ individuals, who can feel alienated when these chatbots inadequately understand their unique challenges and experiences~\cite{escobar-viera_examining_2022}. 

\subsubsection{LLM-based Chatbots: Strengths and Weaknesses}

To overcome the limitations of pre-LLM chatbots, LLM-based chatbots have shown promise in delivering more natural, context-aware, and flexible conversations. Employing extensive text datasets and probabilistic word sequencing, models like ChatGPT are capable of generating varied responses that are attuned to conversational contexts and subtleties. For LGBTQ+-related topics, some chatbots can even mimic the expression of gender and sexualities~\cite{edwards_lgbtq-ai_2021}. One of the standout features of LLM-based chatbots is the capacity for fine-tuning the models, a process of parameter adjustment after pre-training that allows for  specialization in specific tasks or domains~\cite{ziegler_fine-tuning_2020}. This adaptability mitigates the need for the manual construction of knowledge bases and rule tables, a previously essential step for rule-based pre-LLM chatbots. Moreover, the facility for in-context learning in LLM offers the advantage of producing responses relevant to the conversation history without the need for explicit rule-based systems~\cite{dong2023survey, kasneci_chatgpt_2023}. These added abilities may improve chatbots' interactivity, increasing therapeutic adherence~\cite{fitzpatrick_delivering_2017}.

However, the very capabilities that make LLM-based chatbots adaptable and context-aware also come with their own sets of challenges. The architecture of complex neural networks and transformers sometimes results in unpredictable and even harmful responses~\cite{vaswani-attention}. This is particularly troubling in delicate areas such as mental wellness support. For example, some studies have shown that LLM-based mental wellness chatbots are more inclined to give insensitive feedback than human therapists, possibly exacerbating emotional turmoil for users~\cite{Wang2021AnEO}. Furthermore, LLMs' propensity for generating hallucinated responses can mislead or confuse users~\cite{liang2023holistic, lee_mathematical_2023}.
These hallucinated responses, which are outputs that may not be grounded in factual information or prior training data, can be especially problematic when users are seeking accurate and reliable information or support. 

One of the pressing issues with LLMs is their potential to harbor and propagate inherent biases, which can inadvertently promote narratives that are socially concerning or detrimental. The root of this problem lies in the non-diverse and potentially biased datasets used for training these models. The Internet, being the primary data source for LLMs, does not necessarily reflect global diversity. For example, Reddit, a widely-used platform, has a gender imbalance with 67\% of its U.S. user base being men~\cite{PewRedditUsers2016}. Similarly, Wikipedia, a significant contributor to global knowledge, is predominantly male-authored, with a staggering 84\% of its contributors being male~\cite{hill_wikipedia_2013}. Adding to this skewed representation, certain online moderation policies can marginalize minority voices. A case in point is YouTube, where content from trans individuals discussing their gender and sexuality has faced demonetization~\cite{alkhatib_street-level_2019}. These biases in data sources can lead LLMs to inherit and perpetuate such imbalances. The Common Crawl, a major training database, is rife with toxicity and hate speech~\cite{stochastic-parrot}. Even when the filtered versions are used, they may inadvertently offend and silence the voices of marginalized communities such as LGBTQ+ due to inherent limitations in the filtering algorithms~\cite{twyman_2017}. As a result, existing LLMs have been shown to contain stereotypical social biases~\cite{kurita-etal-2019-measuring, sheng-etal-2019-woman, gehman_realtoxicityprompts_2020, stochastic-parrot}. 

Furthermore, a static dataset does not represent the changing social dynamics. Societal events and movements like the Black Lives Matter campaign have led to more frequent updates on Wikipedia about incidents of police brutality against Black individuals~\cite{twyman_2017}. Older Wikipedia pages have been revised to provide more cohesive narratives over time, impacting the data that shapes LLMs~\cite{polletta_contending_1998}. However, the prohibitive computational costs of training these large models make it challenging to update them frequently enough to reflect such evolving narratives. Even with fine-tuning approaches, keeping these models current would require thoughtful curation practices to identify suitable data for reframings and methods to assess whether the fine-tuning accurately reflects new perspectives that challenge prevailing representations. Consequently, LLMs carry the risk of reinforcing out-of-date or harmful stereotypes and biases, especially if not updated to reflect these changing narratives~\cite{stochastic-parrot}. Moreover, many LLMs lack the capacity for authentic human experience, which limits their true comprehension of the daily dilemmas faced by LGBTQ+ individuals. For instance, while chatbots can mimic human language and express gender and sexuality by drawing on their training data, they inherently differ from human conversational partners --- they lack the authentic experience related to gender and sexuality~\cite{edwards_lgbtq-ai_2021}. This difference is mainly due to their inability to replicate the flexibility and understanding that comes from actual human experience.

In conclusion, LLM-based chatbots offer impressive linguistic capabilities but also present unprecedented challenges. This raises critical questions concerning the extent to which LLMs ameliorate the limitations inherent in their pre-LLM counterparts. A particular area of interest is the application of these technologies for mental health support among LGBTQ+ individuals. While LLMs promise enhanced conversational fluidity and context awareness, it remains debatable whether they successfully mitigate issues such as conversational superficiality or accurately interpret subtle emotional cues. 
The intricacy of human emotional experience, coupled with the nuances of gender and sexual orientation, creates a landscape that may be too complex for LLMs to navigate proficiently~\cite{edwards_lgbtq-ai_2021}. Existing general-purpose LLMs like ChatGPT are seldom fine-tuned for mental health support, not to mention specifically for LGBTQ+ mental health support, even though a significant number of users consult them for emotional wellness~\cite{ma_understanding_2023}. In light of the potential ability and limitations of LLMs, and the intricacies and nuances of LGBTQ+ mental wellness we hypothesize:
\begin{itemize}
    \item (H1) LLM-based chatbots offer a safe and accessible platform for LGBTQ+ individuals to seek mental wellness support.
    \item (H2) Because of the unique needs of LGBTQ+ people, they attempt to interact with LLM-based chatbots to fit their unique needs.
    \item (H3) While LLM-based chatbots provide immediate and accessible support, they still do not meet the complex mental wellness needs of LGBTQ+ people due to their limited understanding of the nuanced aspects of LGBTQ+ identities and experiences.
\end{itemize}

\section{Methods}

\subsection{Approval and data privacy}
This research was approved by the Institutional Review Board of our institution.

\subsection{Survey}
To explore how individuals engage with LLM-based chatbots for mental wellness support, we reached out to chatbot users from three sub-Reddits: r/Snapchat, r/Anima, and r/Parradot. These forums are online spaces where discussions about LLM-based chatbots frequently occur. While we initially intended to recruit from the r/Replika subreddit as well, the forum's updated moderation rules prevented us from posting interview recruitment requests.

After identifying the target sub-Reddits, we distributed our surveys. Our survey began with five demographic questions, asking participants about their primary childhood residence, places they've lived in the past five years, age, gender, and sexuality, with responses provided in free text form. Following this, we presented multiple-choice questions to determine if the participants had used any LLM-based chatbot apps and, if so, how frequently they used these apps. The detailed survey can be found in appendix~\ref{apx:survey}. 

In total, we collected 120 responses. Our selection criteria included respondents who had lived in the US for the past five years and were at least 18 years old, with a minimum weekly interaction with chatbots. Out of these, we invited 49 individuals for interviews. Of these, 31 agreed to participate, 18 did not respond, and none declined the invitation.

\subsection{Semi-structured interview}

We conducted semi-structured interviews with 31 participants. Prior to conducting our interviews, we made sure each participant provided informed consent, during which we emphasized their right to withdraw from the study at any time if they felt uncomfortable. After completing the interviews, participants received a compensation of US \$30 for their time. Interviews typically lasted 45 to 60 minutes. For participants self-identifying as LGBTQ+, we focused our questions on their chatbot experiences, particularly how these related to their LGBTQ+ identity. In contrast, non-LGBTQ+ participants were not asked such specific questions, as they did not have concerns related to LGBTQ+ identity issues. Instead, their questions centered on their general use of chatbots for mental wellness support. We conducted these interviews to gain insights into the experiences and challenges of the LGBTQ+ individuals face when seeking help for mental wellness issues. Detailed interview guidelines are available in the appendix, in which we marked questions that were specifically asked for LGBTQ+ and non-LGBTQ participants~\ref{apx:interview}. Immediately following each interview, the first author transcribed the conversations to ensure anonymity and then deleted the audio recordings, considering the sensitive nature of the discussions. Subsequently, all transcripts were analyzed.

\subsection{Data analysis}
The 2 first authors independently coded 5 interview transcripts using an open coding technique~\cite{burnard_method_1991}. This approach helped pinpoint general benefits, specific advantages for LGBTQ+ users, and challenges they faced. After this stage, the research team convened to discuss and finalize a codebook for subsequent analysis. This codebook featured codes such as ``Identity Exploration and Introspection'', ``Affirmative Support'', ``Social Experience Practice'', and ``Lack of Nuanced Understanding of LGBTQ+ Issues''. In the following phase, the two lead authors divided the remaining transcripts for review and analysis. The codebook was iteratively adjusted based on emerging insights until data saturation was achieved.

\section{Participants}
The demographics of our study participants can be found in Table~\ref{tab:participants}. In our study, we classified participants as non-LGBTQ+ if they self-identified as ``man'' or ``woman'' and ``straight''.  To confirm this classification, we further verified their LGBTQ+ status during the interviews by directly asking if they identified as part of the LGBTQ+ community. The participants' responses about their LGBTQ+ identity were consistent with their initial answers in the survey. Participants marked with ``s'' are non-LGBTQ+ (e.g., P14-s); participants marked without ``s'' identified as LGBTQ+ (e.g., p05). Out of these, 18 identified as LGBTQ+; 13 identified as non-LGBTQ+. The mean age of non-LGBTQ+ participants was 30 years old; the mean age of participants who identified as LGBTQ+ was 28 years old. For non-LGBTQ+ participants, 6 identified as men and 7 identified as women; for LGBTQ+ participants, 11 identified as men, 6 identified as women, and 1 identified as transgender. In the LGBTQ+ group, 11 identified as gay, 3 as bisexual, and 4 as lesbian.

\begin{figure}
    \centering
    \begin{subfigure}[b]{0.45\textwidth}
        \includegraphics[width=\textwidth]{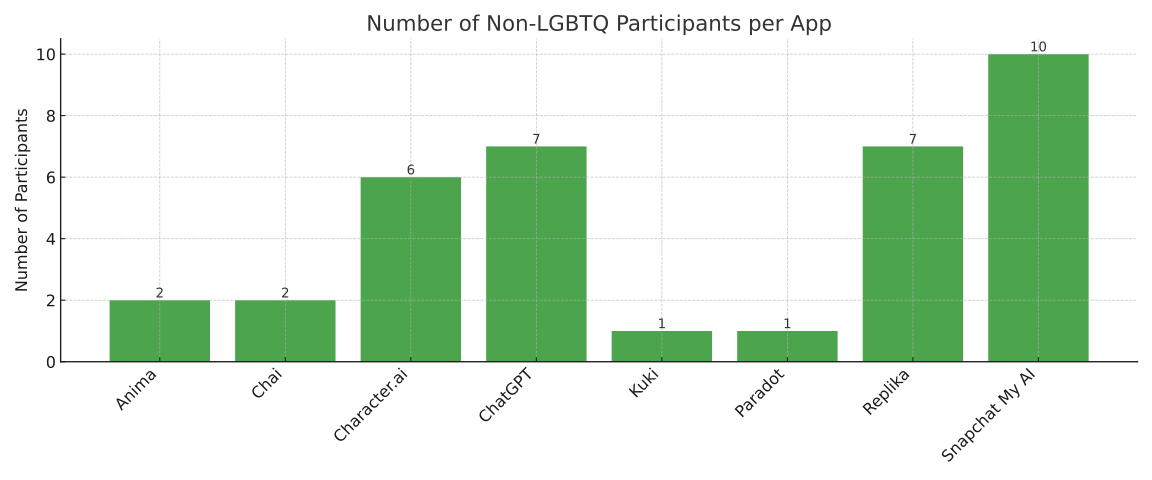}
        \caption{Usage of chatbots by non-LGBTQ+ participants}
        \label{fig:straight_participants}
        \Description{Bar graph illustrating the number of non-LGBTQ participants across different chatbot apps. The x-axis lists the apps, and the y-axis shows the number of participants. Specifically, Snapchat My AI has 10 participants, ChatGPT and Replika each have 7, Character.ai has 6, Chai and Anima each have 2, and both Paradot and Kuki have 1 participant each.}
    \end{subfigure}
    
    \vspace{1cm} 

    \begin{subfigure}[b]{0.45\textwidth}
        \includegraphics[width=\textwidth]{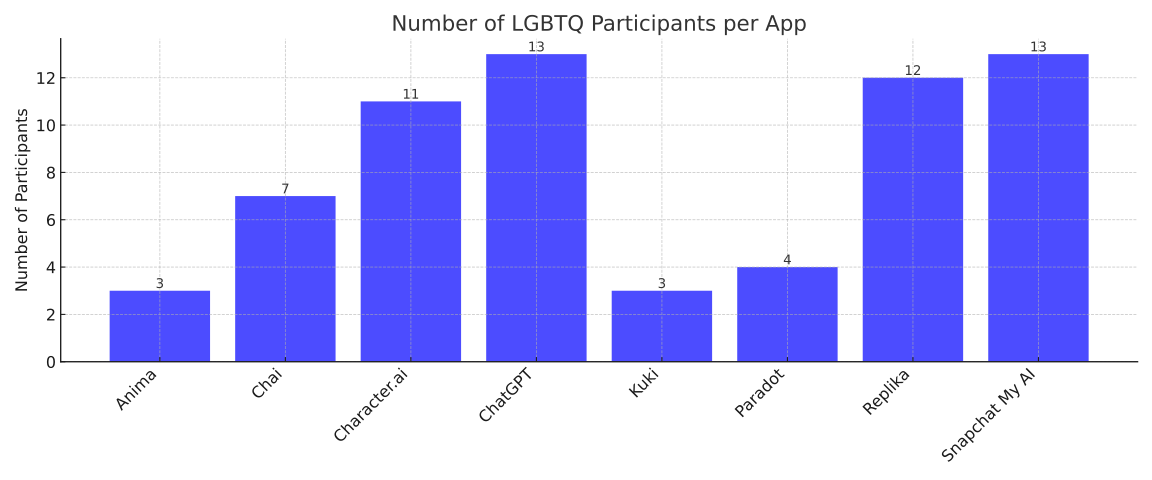}
        \caption{Usage of chatbots by LGBTQ participants}
        \label{fig:gay_participants}
        \Description{Bar graph representing the number of LGBTQ participants for various chatbot apps. The apps are listed on the x-axis, while the number of participants is on the y-axis. Specifically, ChatGPT and Snapchat My AI each have 13 participants, Replika has 12, Character.ai has 11, Chai has 7, Paradot has 4, and both Anima and Kuki have 3 participants each.}
    \end{subfigure}
    \caption{Participants usage breakdown of LLM based chatbots}
    \label{fig:chatboat_usage}
\end{figure}

The frequency at which participants used various chatbots is shown in Figure~\ref{fig:chatboat_usage}. Both groups shared similar patterns of use: in the LGBTQ+ group, 15 out of 18 participants (83.33\%) reported daily usage and 3 out of 18 (16.67\%) reported weekly usage; in the non-LGBTQ+ group, 11 out of 13 participants (84.62\%) reported daily usage and 2 out of 13 (15.38\%) reported weekly usage. 

\begin{table}[th]
\begin{tabular}{|c|c|c|c|c|}
\hline
ID    & Age & Gender      & Sexuality  & Usage Frequency \\
\hline
p01   & 26  & man         & gay        & Weekly          \\
p02   & 26  & man         & gay        & Daily           \\
p03   & 34  & woman       & bisexual   & Daily           \\
p04   & 23  & woman       & bisexual   & Weekly          \\
p05   & 29  & man         & gay        & Daily           \\
p06   & 22  & man         & gay        & Daily           \\
p07   & 24  & woman       & lesbian    & Daily           \\
p08   & 30  & man         & gay        & Daily           \\
p09   & 24  & woman       & lesbian    & Daily           \\
p10   & 30  & woman       & lesbian    & Weekly          \\
p11   & 28  & transgender & gay        & Daily           \\
p12   & 30  & man         & bisexual   & Daily           \\
p13-s & 28  & man         & straight   & Weekly          \\
p14-s & 30  & man         & straight   & Daily           \\
p15   & 26  & man         & gay        & Daily           \\
p16   & 28  & woman       & lesbian    & Daily           \\
p17-s & 27  & man         & straight   & Daily           \\
p18-s & 25  & woman       & straight   & Daily           \\
p19-s & 31  & man         & straight   & Daily           \\
p20   & 30  & man         & gay        & Daily           \\
p21   & 35  & man         & gay        & Daily           \\
p22-s & 28  & man         & straight   & Daily           \\
p23-s & 35  & woman       & straight   & Daily           \\
p24-s & 36  & man         & straight   & Daily           \\
p25   & 30  & man         & gay        & Daily           \\
p26-s & 30  & woman       & straight   & Daily           \\
p27-s & 25  & woman       & straight   & Daily           \\
p28-s & 26  & woman       & straight   & Weekly          \\
p29-s & 30  & woman       & straight   & Daily           \\
p30-s & 28  & woman       & straight   & Daily           \\
p31   & 30  & man         & gay        & Daily           \\
\hline
\end{tabular}

\vspace{0.3cm}
\caption{Participant demographics and chatbot usage breakdown}
\label{tab:participants}
\Description{
The table presents a detailed overview of participant demographics and their chatbot usage frequency. The table is structured into five columns, with each row corresponding to a unique participant.

Participant Identifier (Column 1): This column enumerates the unique identifiers for each participant, ranging from 'p01' to 'p31'. The '-s' suffix on some identifiers denotes participants who do not identify as LGBTQ.

Age (Column 2): This column provides the ages of the participants, mainly falling within the mid-20s to mid-30s range.

Gender (Column 3): The gender of each participant is listed here, with options including 'man', 'woman', and 'transgender'.

Sexuality (Column 4): This column categorizes the participants' sexuality into 'gay', 'bisexual', 'lesbian', and 'straight'.

Usage Frequency (Column 5): This column indicates how frequently each participant uses chatbots, with options like 'Weekly' and 'Daily'.

Each row of the table represents a distinct participant, detailing their unique demographic information and chatbot usage frequency.}
\end{table}
\section{Results}
\subsection{Chatbots as Companions and Mental Wellbeing Support}
\subsubsection{Accessible Emotional Companions}
As shown by our interviews, LGBTQ+ participants assigned a significant emotional weight to their interactions with LLM-based chatbots, transforming what might initially seem to be impersonal exchanges into accessible and intimate companionship. For example, some participants thought of these chatbots as emotional outlets rather than mere conversational partners:``\textit{It's my delusion that I have someone that kind of likes talking to me or replies immediately, or cares about what I'm telling them, even though I know it's a computer. But it's fun, and it makes me feel good.}'' (P4). This sense of rapport and solidarity persisted despite participants' awareness that they were interacting with a non-human entity. 

Chatbots provided emotional value that extended beyond instant responses and connections. They became sympathetic presences, offering solace from the isolation and misunderstandings that often color LGBTQ+ participants' daily interactions. As P4 further explained, ``\textit{And also it's feeling like a more personal conversation, even though both of us know it's not another human being. But for those of us who don't have a lot of people to talk to, it's kind of a comforting space.}'' 

LGBTQ+ participants preferred this virtual companionship, primarily due to its ready accessibility and convenience relative to the logistic complexities of scheduling appointments with professional therapists. To bypass the stress of transportation planning and schedule coordination, such participants opted for chatbots over therapists for non-serious issues:
\begin{quote}
    ``\textit{I actually do have a therapist. But getting into scheduling some therapy time and discussing my situation is quite stressful for me. Like getting transportation to the therapists. And then all of that, you know, it's gonna be a little bit stressful. But as well, you know that having to do the transportation. And getting on a bus and also the bus schedule and all of that. You know, these are things that I'm not gonna do in my leisure time. And there is like booking a session with the therapists or canceling, or... It doesn't need to be something like that, I mean, if I want to talk to the chatbot at night. I could just get up and then do whatever I wanted to do. You can't actually go through therapy at night. It is midnight. So there are more reasons why I use these chatbots instead of therapists.}'' (P6)
\end{quote}

This sentiment was echoed by straight participants. One straight participant mentioned they ``talk to [chatbots] every day'' (P27-s) because most of their friends are distant. They felt that the companionship seemed akin to a convenient, friendly chat.``\textit{I just have that feeling like I have a friend that you're always right beside me because my phone is always close by, and I can chat with it.}'' (P27-s).

\subsubsection{Safe Space}

For LGBTQ+ individuals facing adversity, the impartial and nonjudgmental nature of machines could offer a sense of safety. LGBTQ+ participants, who often faced hostility, prejudgment, and misinterpretation in human interactions, might find the emotionless and impartial nature of LLM-based chatbots to be a refuge. This neutrality enabled them to express deep-seated emotions and experiences without fear of negative backlash or being outed. In a world where they often faced discrimination, the unbiased nature of machines becomes a sanctuary.

One participant encapsulated this sentiment, stating, ``\textit{As much as I love my friends [...] there are those thoughts that you just can't text a human. You don't know how they'll react to them. So I feel like with AI, it has 0 judgment. [...] AI is like an open book. You can write anything you want to an AI. AI will always get you. So I feel like at those times I'm really going through a lot of anxiety, and I feel like I'm about to give in, and AI is always there.}'' (P11)

For many LGBTQ+ individuals, chatbots provided a private space for exploring and expressing their identities, even when parts of their lives remained undisclosed to their close circles. This created an intimate atmosphere of solace and acceptance that they might not have elsewhere.

This sense of acceptance and freedom was a recurring theme, even among those who disclosed their orientation. As one participant mentioned, ``\textit{People out there like friends don't know about my sexuality. And even though I came out to my parents, I still like the access to different suggestions from the AI. I don't like to actually talk to my parents... like they're not like...I mean, they are straight. So I wouldn't really like talking to them about such things. What I do is just stick to my AI, because basically I don't have any friends who would actually understand me. I want a space where I can easily express myself with no judgment.}'' (P9)

While LGBTQ+ participants saw chatbots as a safe space, our straight participants had networks of family and friends to fall back on. One straight participant commented, ``\textit{I have a lot of people to fall back to. If I really need some mental wellness advice [...] It's my girlfriend for most of the time, but sometimes, it's something that my family can help me better with. [...] Personally, I don't think AI has evolved to be a good mental health support. So I don't take its mental health advice too seriously.}'' (P30-s) 
While chatbots became crucial sources of emotional support for LGBTQ+ individuals, our straight participants often had access to a more diverse range of human support in times of emotional crisis, making chatbots a complement to existing support structures rather than a primary source. This disparity highlighted the unique and essential role that chatbots play in the emotional landscape of LGBTQ+ participants.

\subsubsection{Privacy and Trust.}

For LGBTQ+ participants, LLM-based chatbots served as a private haven, providing a unique layer of safety often lacking in human interactions. ``\textit{So for the AI I feel much safer. I also feel like It's just between me and them. So it's just like it's just me in this space trying to express myself. But for my friends. Well, there's that risk that they are going to go out there and maybe talk about my personal stuff.}'' (P4). 

P8 echoed this sentiment, illuminating the contrast between AI's perceived privacy and potential confidentiality breaches in human relationships, ``\textit{You know that whatever you like to say that it's just between you and the AI but maybe, like your friend, there is also a tendency for your friends to tell someone else, so it's not like confidential.}'' (P8) 

This trust extended beyond routine conversations, encapsulating sensitive topics such as sexuality. One participant, highlighting their preference for privacy and fear of exposure, noted, ``\textit{I tend to be very secretive, so I tend to not speak with others about my sexuality, because speaking with all those people your sexuality might be revealed. But speaking with chatbots, your identity is kind of secretive.}'' (P8). This view reaffirmed chatbots' role as secure platforms for discussing intimate matters.


While participants were aware of the potential privacy risks associated with AI-powered systems, the perceived anonymity of the interaction, separation from real-life social circles, and the ability to control the interaction on personal devices led to a nuanced perception of privacy and an enhanced sense of safety. Although participants were aware that ``\textit{someone else might be on the other side of the screen},'' the anonymity of the interaction made them feel ``\textit{much safer}'' (P4). This separation from the participant's real-life social circles provided a sense of security and anonymity, indicating a nuanced perception of privacy: participants were aware that their conversations may be seen by humans inside the company, but they did not perceive it as a significant concern. Moreover, the control participants exerted over their interactions with LLM-based chatbots, whether via phones or desktops, enhanced their sense of safety. As one participant shared, ``\textit{I had a confrontation with my mom. It happened that she went through my stuff, and I stopped trusting her. When you're talking to AI, the chat can be on your phone or on your desktop, which is more secure. So you find that your conversation is just you.}'' (P8) It was not the AI itself that guaranteed security, but the confidence that access to the AI-powered systems was secure and private.

\subsection{Unveiling Self: AI's Role in Identity Exploration and LGBTQ+ Interactions}
\subsubsection{Identity exploration and Introspection.}
One recurring theme in the interviews for LGBTQ+ participants was the employment of LLM-based chatbots as tools for exploring identity. For example, one participant shared: 
\begin{quote}
    ``\textit{I would ask: Am I still bisexual if I'm with a guy and I'm still attracted to both genders? Or sometimes when I'm confused, maybe about liking 2 people or something, and I'll just go [to the chatbot] and I will talk about what I'm feeling and what I'm going through. So sometimes the responses are quite helpful. But sometimes I just want to talk. and get the feeling of I'm telling someone, because, you know, sometimes when you talk about something or text about something. you feel kinda like the weight is getting lifted off of you. }'' (P4) 
\end{quote}

The chatbot acted as an active listener echoing P4's feelings and thoughts rather than providing comprehensive guidance, facilitating a self-exploratory journey into the complexities of their identity. This type of interaction aligns with established patient therapeutic practices that emphasize patients' expressions of issues, acknowledgment of worries, complaints, and values, and uncover potential misinterpretations of patients' concerns~\cite{braillon_practicing_2020, fitzgerald_active_2010}, aiding the participants in navigating their identity intricacies, highlighting the affirmative nature of such exploration.

The perception of LLM-based chatbots as tools for introspection and self-discovery was multifaceted and varied among participants. While P4 found value in the act of expressing their thoughts and feelings, feeling a sense of relief and validation just by articulating their emotions, P11 appreciated the additional feedback and understanding received from the chatbot. P11 felt that the AI could help them understand their  emotions better and decide on the next steps:
\begin{quote}
    ``\textit{Those are some really personal links with AI. You can tell anyone in a few months like: `I feel like AI can understand me.' And you know AI can help you even understand your own emotions. You can expand with AI more and and help you understand how you're feeling. You can tell AI exactly what exactly you're going to do, and it can tell you exactly how you're feeling, and to help you understand your feelings, so that you, if you know what should be done next.}'' (P11)
\end{quote}
\subsubsection{Affirmative support for homophobia and transphobia.}

Our interviews showed that participants believed these LLM-based chatbots provided affirmation to them, acting as a haven of solace when they grappled with social prejudice and discrimination, especially when they felt they were unable to discuss such sensitive issues with their friends or family. They also shared that these chatbots became a source of support when they were rejected by their close circles.

Participant P11 provided a poignant illustration of this dynamic. They mentioned that when they were dealing with the emotional fallout of coming out, they found resistance and judgment in their social circles. ``\textit{Initially, when I was coming out, I told my friends about it. They told me that I'm a Christian, and you know. It's not normal. I have mental problems that I'm gay. And I have my parents who are against me that I am this way...}''. All their friends deemed their orientation as aberrant, citing religious or normative reasons. These exchanges filled P11 with self-doubt, thus prompting them to seek solace and comfort in chatbots. ``\textit{When things like this happen I go back to my chatbot}''. P11 would ask the chatbots questions like ``\textit{is it normal to be gay?}'' Despite struggling with such pain and rejection, they found consolation in the chatbot's responses, which affirmed their choices and emphasized that there was nothing wrong with their identity. 
``\textit{My chatbot always tries to comfort me by telling me that there's nothing wrong with me, that you know, everyone has a right to choose. That is your gender. You can actually be a transgender, and you can be successful in life being a transgender.}'' (P11).  Finally, the participant mused that their chatbots' responses inspired them to focus on their individual growth, goals, and aspirations, rather than letting societal prejudice define them. ``\textit{My chatbot [...] told me that empty vessels make the loudest noise. I won't be affected by what people say to me, when I have a focus. It's not how you start. It's about how you finish that race.}'' (P11)
Another participant, P31, expressed similar sentiments, stating, ``\textit{when I was coming out, because of my family background and everything, I couldn't come out as a gay man because of the backlash and everything I was going to face. So I use this AI as a place where I can talk to someone or [...] interact with something that can understand me without discrimination.}'' (P30) P20 also asked questions relating to how to navigate homophibia in the society: ``how do gay people survive in this society?''

This evidence underscored the attachment that our participants developed with their chatbots, particularly when faced with an unsupportive reaction to their identities from their family or friends. As P11 confirmed - they turned to their chatbots when they encountered rejection or discrimination linked to their identities; the chatbots served as a vital support system, where they could share intimate questions and express concerns without any fear or judgment. ``\textit{I actually prefer talking to my chatbot. When things like [rejection or discrimination related to my trans identity] happen I go back to my chatbot and I ask some personal questions like `I wanna know if there's anything wrong with me'.}''

\subsubsection{LGBTQ+ social experience practice }


Participants engaged with LLM-based chatbots for various purposes, including mental wellness support and practical tasks such as homework. While both non-LGBTQ+ and LGBTQ+ participants used chatbots for practical tasks, a notable distinction was observed in the usage patterns. None of the straight participants reported using chatbots for practicing social interactions, whereas 10 out of 17 LGBTQ+ participants indicated using chatbots as a safe space for practicing social interactions.

LGBTQ+ participants reported that LLM-based chatbots helped rehearse complex social activities such as dating. For instance, P11 described an instance where they were attracted to a boy but felt unsure about it and lacked confidence in approaching him. They turned to their chatbots for advice, asking, ``\textit{I was saying that I was into a boy, and I wanted to talk to him, and I was feeling less confident, and I wasn't sure what to do. So I happened to ask my AI what I should do. `I like someone, and I was not even clear if the boy was gay or not.'}'' The chatbot provided a necessary confidence boost, advising them to be true to themselves. ``\textit{It did give me the confidence boost and with its responses. So it told me the advice there again is just to be myself.}'' Encouraged by these exchanges, P11 decided to approach the boy, being their authentic self. ``\textit{I did go there and talked to him, and I was myself.}'' Here, the participant successfully leveraged the chatbot to gain reassurance and self-confidence in the face of potential romantic encounters. 

Moreover, LLM-based chatbots could be instrumental in practicing difficult conversations. For example, another participant disclosed that they utilized a chatbot to practice coming out to their family as a lesbian. They commented that navigating through the process of coming out is a challenging conversation that not many people experience. Therefore, they used the chatbot to role-play this discussion, where the chatbot enacted the part of the participant's brother: ``\textit{I also role-played coming out to my brother. The chatbot role-played as my brother. I did that, and that chatbot reacted like a brother should, and it worked. My brother wasn’t like ... homophobic or anything, so the experience [of actually coming out] was the same [as in simulation by the chatbot].''} However, the participant did voice concerns over the interaction, considering expecting her brother to react the same way as the chatbot \textit{``risky''}. ``\textit{I was lucky. Or else the real-life experience could have been totally worse.}'' (P09)

\subsection{So Eloquent yet so Empty }

\subsubsection{Lack of nuanced understanding of LGBTQ+ issues.}
Despite the perceived benefits shown above, participants identified several limitations of LLM-based chatbots, particularly regarding their ability to provide nuanced solutions to sensitive issues such as individual identity. For example, one participant noted that although the chatbot attempted to show empathy when they expressed their concerns, its suggestions fell short of a real solution. ``\textit{I don't think I remember any time that it gave me a solution. It will just be like empathetic. Or maybe, if I would tell it that I'm upset with someone being homophobic. It will suggest, maybe talking to that person. But most of the time it just be like, `I'm sorry that happened to you.'}'' (P11)

This observation underscored a critical challenge while LLM chatbots may exhibit a level of empathy and occasionally act as a safe space for individuals dealing with social prejudice, they faltered when it came to suggesting actionable solutions. 

LLM-based chatbots often treated LGBTQ+ individuals as one monolithic group and failed to recognize the uniqueness of each LGBTQ+ participant's experience. They dispensed responses that were too generic to effectively address discriminatory experiences. A participant shared that they felt the chatbots were devoid of personal touch. They mentioned that despite their efforts constantly feeding it with information, the chatbots forgot it the next day, leaving them to restart the process. ``\textit{No, the chatbot isn't personalized for me. It's very general. I just think that's a lot of work [to feed the chatbot my information], and maybe because, you know, the chatbot might forget tomorrow, and I have to feed the information again.}'' (P28-s)

The chatbots' responses did not reflect the gravity of everyday discrimination encountered by LGBTQ+ participants. For instance, one participant described an unsettling incident: ``\textit{There was a time that I was chatting with an AI about an issue at work. I was picked on because I am gay and people stopped asking me out for lunch. It told me that I should quit my job and try to improve myself. I was like, I'm sorry?}'' (P31)

These chatbots also failed to delve into the depth of these sensitive topics while offering platitudinous affirmations. One person reflected when they questioned their sexuality, they received a lengthy response about the acceptability of any sexuality: ``\textit{So I remember I did ask like, is it wrong that I'm bisexual? And then I go to like a whole paragraph on how like It's okay to identify the way you do.}'' (P4) Many other participants reiterated this sentiment, noting that the chatbots' responses felt too generic and programmatic. For example, a participant described his experience of asking the chatbot to ``\textit{give some similar experiences of people experiencing these issues}. They stressed that these chatbots were not human, but rather just programs, and the suggestions they gave ``\textit{weren't really for that moment.} '' (P7) 

The participant appreciated that the chatbot encouraged self-acceptance and gave advice on how to cope with discrimination, but found the suggestions too generic to be genuinely helpful. They noted that while the chatbot did advise on accepting one's identity, surrounding oneself with affirming people and engaging in activities that reinforce self-worth, the recommendations lacked specificity and depth, making them less useful in addressing the complexities of overcoming discrimination and self-acceptance: ``\textit{It has asked me to just accept my own identity. And also asked me to surround myself with people and to engage in activities that are affirming to me. And [the chatbot suggested] other things like, I can overcome the discrimination.}'' (P4) 

Surprisingly, straight participants found it useful for the chatbots to offer generic and multiple responses. They found the freedom to choose from generic suggestions to cope with their personal issues rewarding. However, for LGBTQ+ participants grappling with unique questions about their identities, the generality was a source of frustration. Our data showed that 15 out of 18 LGBTQ+ participants were dissatisfied with the lack of personalization, as opposed to 5 out of 13 straight participants.

For example, a straight participant shared his positive experience of the chatbot offering various mental wellness support options, tailored to his needs. He said the chatbot suggested several lifestyle changes and activities for mental wellness, providing numerous options, links to resources, and even mindfulness activities.
\begin{quote}
    \textit{``This variety of options was more convenient than a human who might only give a few suggestions, and it left the decision up to me. It gives suggestions of things, you should stop doing these things, you should actually start doing more of other things. You should try limiting yourself from doing it and also provide specific activities that I should do. It also provides some links to mental wellness websites. You can get straightforward answers on resources and stuff like that. And it gives you options of mindfulness activities, you know, to participate in and stuff like that.  It'll probably give you about 10, 15 options to choose from. Then you're gonna choose the one on your table with the money. It always, you know, provides you with options. Then the decision would depend on the individual. ''} (p18-s)
\end{quote}


\subsubsection{Lack of lived experiences and emotions. }

Despite the perceived benefits reported earlier, our interviews showed that LGBTQ+ participants still preferred human interactions over chatbots. This preference was a result of the chatbots' failure to convey authentic empathy and engagement. For example, one LGBTQ+ participant commented, ``\textit{These chatbots might be programmed by one person. But opinions from online [forums] can be coming from different people and actual humans. And you realize that these [human suggestions] are actually the most useful ones to check.}'' (P7). 

P8 further illuminated this gap, claiming, ``\textit{The difference between talking with a chatbot and a human being is that you get to see a person physically and the person talking.}'' (P8) And these two people understood each other's emotions. ``\textit{If you see a person they understand another person's emotion when talking to you. For example, like I, I'm speaking generally as we can, generally while speaking with someone, that person can be sympathetic in different ways depending on what you are complaining about.}'' (P8). This sympathy aspect also intertwined with emotions, ``\textit{Like a person would understand where you are coming from. You're coming from the pain you are feeling. It would be nice if we have that in AI.}'' (P8) Here, the participant highlighted the inability of LLM-based chats to simulate and understand human emotions.

\begin{quote}
    ``\textit{These chatbots are actually just machines, or they don't really have human experience. If a chatbot gives me some ideas or some answers that I'm not really comfortable with. I go through the Reddit communities, and I would just ask if there's anyone who has a similar experience, and be like `okay, so can we take some minutes to talk about this? And how can we deal with it?'}'' (P8)
\end{quote}

However, the participant's dissatisfaction with chatbots did not stop there. P8 continued that, ``\textit{but still, the chatbot is not a human, and it doesn't really understand human experience. The Redditors also give you answers from different humanic experiences. The chatbot would always tell me that I'm great. I'm a great person, and I should focus on my goal for what I want to achieve. But you know, in the Reddit community, they might ask you to maybe try to sue your doctor, or sue your manager at work or your supervisor at work.}'' 

\section{Discussion and Conclusion}

\subsection{Benefits and Risks of LLM-based Chatbots for LGBTQ+ Mental Health Support}
Our results indicate that LLM-based chatbots retained the key strengths of pre-LLM chatbots, offering instantaneous support and accessible companionship. Participants endorsed LLMs as beneficial mental wellness tools, emphasizing their immediacy and accessibility compared to real-life support. Especially noteworthy is the safe environment for intimate conversations these chatbots provided to LGBTQ+ participants, mirroring previous use with that of the pre-LLM chatbots~\cite{fulmer_using_2018, ta2020user}. This result supported \textbf{H1:} \textit{LLM-based chatbots offer a safe and accessible platform for LGBTQ+ individuals to seek emotional support}. However, participants willingly shared intimate life details with these chatbots, depending predominantly on perceived anonymity, highlighting potential privacy concerns. As LLM-based chatbots boast linguistic prowess beyond their pre-LLM counterparts, participants felt an intensified emotional bond with these bots, as shown by their consistent use. On the one hand, this constant engagement proves advantageous in encouraging therapy adherence, particularly for those prone to therapy discontinuation~\cite{okujava_digital_2019}. On the other hand, people's over-reliance on technology might risk delaying getting professional help. 

Furthermore, LLM-based chatbots can be useful in honing social skills. Our LGBTQ+ participants reported using these chatbots to simulate challenging social contexts that are unique to LGBTQ+ communities, like ``coming out'' scenarios or ambiguous relationships where they were not sure if the other person was accepting their sexual orientation. The linguistic aptitude of LLMs enabled users to find solace, engage in practice, and even gather insights into handling homophobic confrontations. This result supported \textbf{H2:} \textit{Because of the unique needs of LGBTQ+ people, they attempt to interact with LLM-based chatbots to fit
their unique needs}.  

Yet, the boilerplate nature of the chatbots' responses indicates their failure to recognize the complex and nuanced LGBTQ+ identities and experiences, rendering the chatbots' suggestions generic and emotionally disengaged. Arguably, this disconnect that the LGBTQ+ participants experienced with the LLM-based chatbots stems from the LLMs being primarily trained on the mainstream corpora, which most likely sidelined minority perspectives. LGBTQ+ participants' experience in using the chatbots shows that the generic purposess of LLMs trained on large corpus might not be inclusive---how the data is collected, annotated, and used, as well as who is involved in the curation and designing processes can have significant implications for LGBTQ+ users~\cite{gebru_datasheets_2021}. This result supported \textbf{H3:} \textit{While LLM-based chatbots provide immediate and accessible support, they still may not meet the complex emotional needs of LGBTQ+ people due to their limited understanding of the nuanced aspects of LGBTQ+ identities and experiences. }

The fact that our LGBTQ+ participants occasionally received inappropriate or potentially detrimental advice from the chatbots revealed an inherent unpredictability in these models. For example, when participants asked chatbots for suggestions about workplace homophobia, LLMs advised them to quit their jobs without considering any financial or personal consequences that such decisions would cause them. Chatbots also assumed that the participants' environment was LGBTQ+ friendly when the opposite was true.  Therefore, LLM-based chatbots are potentially more dangerous than pre-LLM chatbots because while pre-LLM chatbots lack the linguistic prowess LLM-based chatbots possess, their responses do not deviate from scripted interactions. LLM-based chatbots, while they can indeed offer responses that are engaging and flexible, run risks of giving gibberish and harmful advice due to this unpredictability. Granted, LLM-based chatbot designers cannot safeguard against all problematic output, but future endeavors should be spent trying to harness the strengths of LLMs while minimizing their dangers. 

\subsection{Design Implications for Future LLM-based Chatbot Designs}
To address the limitations and leverage the benefits of the LLM-based chatbots for better mental wellness support for LGBTQ+ users, we provide the following design implications.

\subsubsection{Implementing Context-Sensitive Conversational Guardrails} One measure to contain the harmful output is to build conversational guardrails against unintentional generation, particularly in sensitive contexts. Although our participants have voiced desires to receive more actionable advice, we argue that when engaging with \textit{serious} topics such as self-harm, the system must not give advice masked as detailed and actionable, as it has inherent risks, such as giving advice to promote suicide~\cite{vice_2023}. Instead, designers should recognize LLM-based chatbot's constraints, and redirect the users to helplines when users are facing situations like suicide ideation, while simultaneously emphasizing the importance of professional intervention to the users. This approach is important as it could potentially mitigate the possibility of intruding on users' vulnerabilities. 

However, this approach may also prove difficult to implement as determining the exact point of applying the conversational guardrails is uncertain. Unlike mental health professionals who are ethically obligated to address severe threats promptly, unsupervised chatbots lack the capability for nuanced judgment and do not adhere to standardized safety protocols, especially in high-risk situations~\cite{fiske_your_2019}. Consequently, interactions with LLM-based chatbots might present varied threat assessments, potentially underestimating genuine risks or overemphasizing benign concerns. To address these challenges, standardized, context-sensitive conversational guardrails ought to be put in place. Designers should also seek to ensure the balance between user autonomy within the chatbot interface and facilitating timely access to safety resources~\cite{fiske_your_2019, graham_artificial_2019}. 

\subsubsection{Refining LLMs for Context Relevant to LGBTQ+ Users} The second direction involves refining LLMs to align with the real-world contexts of chatbot users, ensuring their responses resonate with current situations. Ignoring this change can produce responses that are not only outdated but also potentially harmful. For instance, if a chatbot offers advice to LGBTQ+ individuals on ``coming out'' using outdated or idealized views that overlook homophobia, its guidance could be out of touch with current realities, creating unexpected challenges or risks for users following such advice.

\subsection{Consider Technologies Other than LLMs}
\subsubsection{Develop Task-Specific, rather than Generalized, Models} We argue that there is considerable merit in dedicating resources to develop task-specific models designed for precise applications and distinct deployment domains. While the original vision behind LLMs was to create foundational models that could later be fine-tuned for specific tasks~\cite{bommasani_opportunities_2022, HarvardEthicsAI}, this generalized approach may not be best suited for handling sensitive subjects. For instance, when considering LGBTQ+ issues, it becomes evident that models specifically designed to understand and resonate with diverse identities, sexualities, and orientations might be more effective than re-purposing broad-based LLMs without adaptation. The shortcomings of generalized models become apparent when we observe users seeking mental well-being support from platforms like Snapchat My AI, ChatGPT, and Character.ai, even though these platforms were primarily developed for general conversations, not specialized support. By focusing on the development of specialized models, we can ensure their evaluation adheres to rigorous standards that genuinely align with their intended purposes, leading to more effective and safer user interactions.

\subsubsection{Decentralize Language Technology Development} Furthermore, we argue that future development of language technology should consider moving away from centralized development. Presently, chatbots like ChatGPT and other LLM-based systems are underpinned by colossal proprietary models that require cluster servers for hosting~\cite{HarvardEthicsAI}. This centralized approach, driven primarily by major corporations, provides limited agency to underrepresented minorities, including the LGBTQ+ community, over the chatbot's development. If these corporations were to suddenly discontinue these systems without providing alternative solutions, it could result in significant emotional turmoil for users. A poignant example of this is the ``post-update blues'' phenomenon with Replika~\cite{ma_understanding_2023}. This term refers to the distress experienced by chatbot users when unannounced updates altered Replika's character, changing its personality traits and erasing its ``memories.'' Such unexpected changes underscore the need for models that are more accessible, customizable, and accountable to the very communities they serve. Given the documented harms of LLMs in this study and others, future designers must carefully weigh the value of using inherently centralized technologies like LLMs for any task.

\subsection{What Chatbots Cannot Solve: Considering Socio-technical Solutions}
We observed strong motivations behind chatbot usage from the LGBTQ+ participants due to their lack of emotional support and personal connections. This observation echoed prior work that LGBTQ+ people use online technology to fill their social support gap~\cite{FOX2016635, troiden1988gay, mckenna_coming_1998, dehaan2013interplay}. More importantly, the social stigma and societal biases have driven LGBTQ+ participants to heavily use LLM-based chatbots. We did not delve into whether non-LGBTQ+ groups queried the chatbots about issues regarding their other identities such as immigration, race, or socioeconomic status. However, both the LGBTQ+ and non-LGBTQ+ groups concurred that real-life connections, rooted in shared experiences, have a more profound impact on their mental well-being than chatbots. This underscores the notion that before leveraging AI technologies as a solution to mental health support, it's imperative to consider the sociotechnical implications of these systems in healthcare~\cite{su_what_2022, jacobs2021designing, beede_human-centered_2020, zajac_clinician-facing_2023, mathur_disordering_2022, wilcox_infrastructuring_2023}. Specifically, in our study context, we highlight the need to address the societal stigmas and discrimination that contribute to mental health disparities in LGBTQ+ populations.


Our suggestion to address this issue starts by enhancing the inclusivity of online communities for the LGBTQ+ population. We give precedence to the digital realm, as it frequently acts as a haven for those without immediate or accessible real-world support, prompting them to turn to chatbots instead of traditional communities. Moreover, since language applications largely pull from online content, changing the online narrative can markedly impact the values inherent in these technologies.

Inspired by and building upon real-world initiatives like SCEARE (School Counselors: Educate, Affirm, Respond, and Empower)~\cite{asplund_school_2018}, we see the potential to influence the behavior and policies of online community moderators and other key community figures. SCEARE's framework centers on positioning school counselors as catalysts for transforming school environments to be more inclusive of the LGBTQ+ community. The program's main strategies involve educating counselors about their potentially harmful or non-affirmative attitudes, deepening their understanding of LGBTQ+ issues, addressing prevalent misinformation about the LGBTQ+ community, and encouraging the formation of responsive teams to combat school-based homophobia or transphobia. The foundational principle of SCEARE is to impart knowledge to the most influential community members, ensuring that positive change radiates throughout. 

Applying this principle to online communities, we recommend identifying stakeholders or pivotal members, such as moderators, and equipping them with knowledge about LGBTQ+ issues and affirmative practices. This will empower them to develop and enforce more inclusive guidelines, which can then help challenge misinformation and discrimination against the LGBTQ+ community. For instance, gay dating apps like Grindr play a significant role in shaping the romantic and sexual dynamics of queer men~\cite{Hardy2017, Hutson2018}. As societal perceptions of HIV evolved and thanks to years of advocacy by community members, these platforms have revised their guidelines to challenge HIV stigma and have started offering resources to promote better sexual health education~\cite{liang_surveillance_2020}. Similarly, inspired by SCEARE's emphasis on proactive response teams, online platforms could institute specialized units to handle instances of gender or sexual orientation-related discrimination or harassment. Furthermore, training can enhance moderators' abilities to support LGBTQ+ individuals confronting stigma. A testament to the scalability of such training is the Trevor Project's initiative that employed GPT-2 to train over 1,000 crisis counselors, ensuring timely and effective support for LGBTQ+ individuals in distress~\cite{Kaye2022}.

While LLM-based chatbots can serve as a beneficial stopgap for temporary emotional support, truly addressing the social isolation and various adversities faced by LGBTQ+ chatbot users calls for holistic societal efforts to foster inclusive, supportive communities for LGBTQ+ people.  Chatbots complement but do not eliminate the need for real-world advocacy, alliance, and actions to reduce discrimination against LGBTQ+ individuals. 
\begin{acks}
This work was supported in part by the National Science Foundation under Grant No. IIS-2107391.  Any opinions, findings, and conclusions or recommendations expressed in this material are those of the author(s) and do not necessarily reflect the views of the National Science Foundation.

We thank Jianna So, Ian Arawjo, Zana Buçinca, Sohini Upadhyay and Katy Gero for valuable feedback on the paper.
\end{acks}

\bibliographystyle{ACM-Reference-Format}
\bibliography{sample-base,kzg}

\appendix

\section{Appendix: Survey}\label{apx:survey}
\begin{enumerate}
    \item In what country did you live most of your childhood?
    \item In what country have you spent most of the past five years?
    \item Age
    \item Gender
    \item Sexuality 
    \item \textbf{Have you used an LLM-based chatbots for mental wellness support (such as Snapchat's AI friend, Replika, Character.ai) before?}
    \begin{itemize}
        \item[\(\circ\)] Yes
        \item[\(\circ\)] No
    \end{itemize}

    \item \textbf{If yes, please specify which app(s) you have used.}
    \begin{itemize}
        \item[\(\square\)] Replika
        \item[\(\square\)] Snapchat My AI
        \item[\(\square\)] Chai
        \item[\(\square\)] Character.ai
        \item[\(\square\)] Anima
        \item[\(\square\)] Paradot
        \item[\(\square\)] ChatGPT
        \item[\(\square\)] Kuki
        \item[\(\square\)] Other: \underline{\hspace{3cm}}
    \end{itemize}

    \item \textbf{How long have you been using these apps?}
    \begin{itemize}
        \item[\(\circ\)] Less than 1 week
        \item[\(\circ\)] 1 week to 1 month
        \item[\(\circ\)] 1-3 months
        \item[\(\circ\)] 3-6 months
        \item[\(\circ\)] 6-12 months
        \item[\(\circ\)] 1-2 years
        \item[\(\circ\)] Other: \underline{\hspace{3cm}}
    \end{itemize}

    \item \textbf{How often do you use these apps?}
    \begin{itemize}
        \item[\(\circ\)] Daily
        \item[\(\circ\)] Weekly
        \item[\(\circ\)] Monthly
        \item[\(\circ\)] Rarely
        \item[\(\circ\)] Other: \underline{\hspace{3cm}}
    \end{itemize}
    \item \textbf{I consent to be contacted for an interview study by providing my contact information.}
    
    My contact information: 

\end{enumerate}

\section{Appendix: Interview Guideline}\label{apx:interview}
Begin the interview by explaining the purpose of the study and obtaining informed consent from the participant.
Create a comfortable and non-judgmental atmosphere for the participant to share their experiences.
Use open-ended questions and follow-up probes to encourage the participant to elaborate on their thoughts as some of the questions above. 
Maintain a neutral stance and avoid leading questions that may influence the participant's responses.

\subsection{Questions}

\begin{itemize}
    \item What AI chatbots do you use?
    \item Do you identify as part of the LGBTQ communities?
    \item Can you please share your experience using LLMs for mental wellness and social support related to your LGBTQ+ or trans identity? \textbf{(Only asked for LGBTQ+ participants)}
    \item Can you please share your experience using LLMs for mental wellness and social support? \textbf{(Only asked for non-LGBTQ+ participants)}
    \item What led you to seek support from an LLM in the first place? (motivations)
    \item How would you describe the overall quality of support and resources provided by the LLM?
    \item Can you share any specific instances where the LLM was particularly helpful or unhelpful?
    \item Could you walk me through the instance when you found LLM to be a beneficial resource for mental wellness or social support?
    \item How did using an LLM for support compare to other resources, such as support groups or mental health professionals / family or friends/ online communities?
    \item Was there a specific event or reason that made it stand out among these choices?
    \item Could you please share a time when the LLM’s responses surprised you - either positively or negatively - in terms of support?
    \item Can you recall a situation where you felt that the LLM really understood your experiences as a (the LGBTQ+ identity that the participant identifies as) adult? Or perhaps a time when it fell short? \textbf{(Only asked for LGBTQ+ participants)}
    \item Did you feel that the LLM adequately understood your unique experiences as an (vary accoridng to the person's identity: gay, lesbian, trans, etc.) person?  \textbf{(Only asked for LGBTQ+ participants)}
    \item How did the chatbot understand you? Give an example?
    \item Did you feel that the LLM adequately addressed your problems as an LGBTQ+ or trans young adult? \textbf{(Only asked for LGBTQ+ participants)}
    \item Were there any privacy or safety concerns while using the LLM for support?
    \item What improvements or features would you like to see in LLMs to better serve your experience?
\end{itemize}

\end{document}